\documentclass[prc,showpacs,showkeys,superscriptaddress,nofootinbib,twocolumn,floatfix]{revtex4}
\usepackage{graphicx,color,mathrsfs,amsmath,amssymb,amsthm,amsopn,bm}
\usepackage[nottoc]{tocbibind}
\usepackage[normalem]{ulem}
\newlength{\lslashl}

 \renewcommand\sout{\bgroup \color{red} \ULdepth=-.5ex \ULset}

\begin{document}

\title{\boldmath
Medium Modifications of the $\rho$ Meson in Nuclear Photoproduction}

\author{F. Riek}\email{friek@comp.tamu.edu}
\affiliation{Cyclotron Institute and Physics Department,
Texas A\&M University, College Station, Texas 77843-3366, USA}

\author{R. Rapp}\email{rapp@comp.tamu.edu}
\affiliation{Cyclotron Institute and Physics Department,
Texas A\&M University, College Station, Texas 77843-3366, USA}

\author{Yongseok Oh}
\affiliation{School of Physics and Energy Sciences, Kyungpook National
University, Daegu 702-701, Korea}

\author{T.-S. H. Lee}\email{lee@phy.anl.gov}
\affiliation{Physics Division, Argonne National Laboratory, Argonne,
Illinois 60439, USA}

\begin{abstract}
We extend our recent study of dilepton invariant-mass spectra from
the decays of $\rho$ mesons produced by photon reactions off nuclei.
We specifically focus on experimental spectra as recently measured by
the CLAS Collaboration at the Thomas Jefferson National Accelerator
Facility using carbon and iron nuclei. Building on our earlier work,
we broaden our description to a larger set of observables
to identify sensitivities to the medium effects predicted by
microscopic calculations of the $\rho$ spectral function.
We compute mass spectra for several target nuclei and study the
spectral shape as a function of the 3-momentum of the outgoing
lepton pair. We also compute the so-called nuclear transparency
ratio which provides an alternative means (and thus consistency check)
of estimating the $\rho$ width in the cold nuclear medium.
\end{abstract}

\date{\today}
\maketitle

\section{Introduction}

The study of in-medium properties of vector mesons has been pursued
vigorously in recent years. The restoration of the spontaneously broken
chiral symmetry at high temperatures and/or density predicts that hadron
properties change when approaching pertinent phase changes: the spectral
functions of chiral partners are believed to
degenerate~\cite{Rapp:1999ej,Hayano:2008vn,Rapp:2009yu,Leupold:2009kz}.
Precursor effects of chiral restoration may already occur in more dilute
hadronic matter where medium modifications can be computed in more
controlled calculations using many-body techniques. How far such
medium effects can be related to fundamental properties of the QCD
phase transition(s) remains an open question.
On the experimental side, electromagnetic probes (real and virtual
photons) are very suitable because no significant distortions occur owing to
strong initial and/or final-state interactions with the nuclear medium. 
Invariant-mass spectra of dileptons in heavy-ion collisions
(HICs) at the Super Proton Synchrotron (SPS)~\cite{Arnaldi:2006jq,Adamova:2006nu,Tserruya:2009zt}
clearly revealed the presence of medium modifications of the $\rho$-meson
spectral function in the hot and dense medium. The most recent NA60 dimuon
data~\cite{Arnaldi:2008er} enable {\em average} medium effect
on the $\rho$ meson to be quantified as an approximately 3-fold broadening relative to its free
width with little (if any) mass shift~\cite{vanHees:2007th,Rapp:1999us}.
This average, however, results from a time evolution of a hot and dense
fireball encompassing hadron densities which vary by approximately a factor
of 5 from the critical temperature ($T_c\simeq175$~MeV) down to
thermal freezeout ($T_{\rm fo}\simeq120$--$130$~MeV). Thus, the predictive power
under such conditions requires not only a good knowledge of the time evolution
of temperature and baryon density, but also a well-constrained model for
the in-medium physics figuring into the vector-meson spectral functions.
More elementary reactions using a single-particle projectile directed
on nuclear targets offer a valuable simplification of both aspects:
(a) the nucleus provides a static medium (at zero temperature) and
(b) the medium consists of nucleons only (rather than including a tower
of mesonic and baryonic excitations). A further benefit is that medium
effects induced by nucleons (baryons) were indeed identified as the
more relevant one relative to
mesons~\cite{Rapp:1999us,Cabrera:2000dx,Eletsky:2001bb,Post:2000qi}.
Although these features should, in
principle, allow for a better theoretical control in the evaluation of
medium effects, there is, of course, a price to pay:
(i) the medium density is restricted to that of normal nuclear matter,
$\varrho_0^{}=0.16$~fm$^{-3}$, with significant density gradients
owing to surface effects, and (ii) the production mechanism is more involved
(generally depending on the projectile) compared to an approximately
thermal medium in ultrarelativistic HICs.
Nevertheless, if the latter point can be addressed satisfactorily (e.g.,
by gauging the production process on proton targets), valuable
information on medium effects is expected to emerge from the investigation
of nuclear dilepton production experiments.

Dilepton spectra in elementary reactions were measured using both
proton- and photon-induced production off nuclei. Using a 12~GeV proton
beam at KEK, the E325 Collaboration reported a significant reduction in
the $\rho$ mass~\cite{Naruki:2005kd}. However, with a
$1$--$3.5$~GeV incident-energy photon beam at JLab, the CLAS
Collaboration~\cite{Clas:2007mga,Wood:2008ee}, using an absolutely
normalized background subtraction procedure, found not
a significant mass shift but rather a moderate broadening of the
dilepton excess spectrum associated with $\rho$ decays (cf.~also
Ref.~\cite{Huber:2003pu} for two-pion production experiments on light
nuclei).

As indicated above, the starting point for a theoretical description
of these reactions is a realistic model for the elementary production
process, $\gamma\,N\,\rightarrow\,e^+e^-N$ as studied by several
authors~\cite{Schafer:1994vr,Effenberger:1999ay,Oh:2003aw,Lutz:2005yv}.
In our previous study~\cite{Riek:2008ct}, we combined the
meson-exchange model of Ref.~\cite{Oh:2003aw} with an in-medium
$\rho$ propagator~\cite{Rapp:1999us} which describes low-mass dilepton
and photon spectra in HICs at the CERN-SPS~\cite{vanHees:2007th}.
The production amplitude was augmented by baryon resonance channels
to ensure consistency with the medium effects in the $\rho$ propagator.
Using an average local-density approximation for the decay points of
the $\rho$, this model was found to describe the dilepton invariant-mass
spectra off iron targets, as measured by the CLAS Collaboration, fairly
well.

In this article, we improve and extend our previous
work~\cite{Riek:2008ct}
in several respects. First, we refine our schematic model for estimating
the density probed by the $\rho$ meson by folding over a realistic density
distribution (rather than using an average density), thereby accounting for
momentum and density dependence of the decay rate.
The improved description is compared to existing carbon and iron data,
supplemented by predictions for several heavier targets to illustrate the
sensitivity to medium effects when going up to uranium.
We furthermore investigate the role of 3-momentum cuts on the outgoing
lepton pair as an additional means to enhance the observable medium
modifications. Finally, we compute the so-called nuclear transparency
ratio as an independent (and thus complementary) observable to obtain
quantitative information on the in-medium width of the $\rho$ meson, by
measuring how the absolute magnitude of the cross section depends on
the nuclear mass number.

This paper is organized along the aforementioned lines. In Sec.~\ref{sec_pro-dec}
we briefly recall the main ingredients of our approach, the improvements
and the resulting density profiles probed by the $\rho$ decays.
In Sec.~\ref{sec_signatures} we update the comparison to existing data
(Sec.~\ref{ssec_update}) and carry out the sensitivity studies with
respect to the target (Sec.~\ref{ssec_target}) and 3-momentum
(Sec.~\ref{ssec_mom}) dependence of the spectra, as well as the
transparency ratio of the total yields (Sec.~\ref{ssec_transp}).
We conclude in Section~\ref{sec_concl}.

\section{\boldmath $\rho$ Production and Decay in Nuclei
\label{sec_pro-dec}}
%
\begin{figure*}[t]
\includegraphics[scale=0.7]{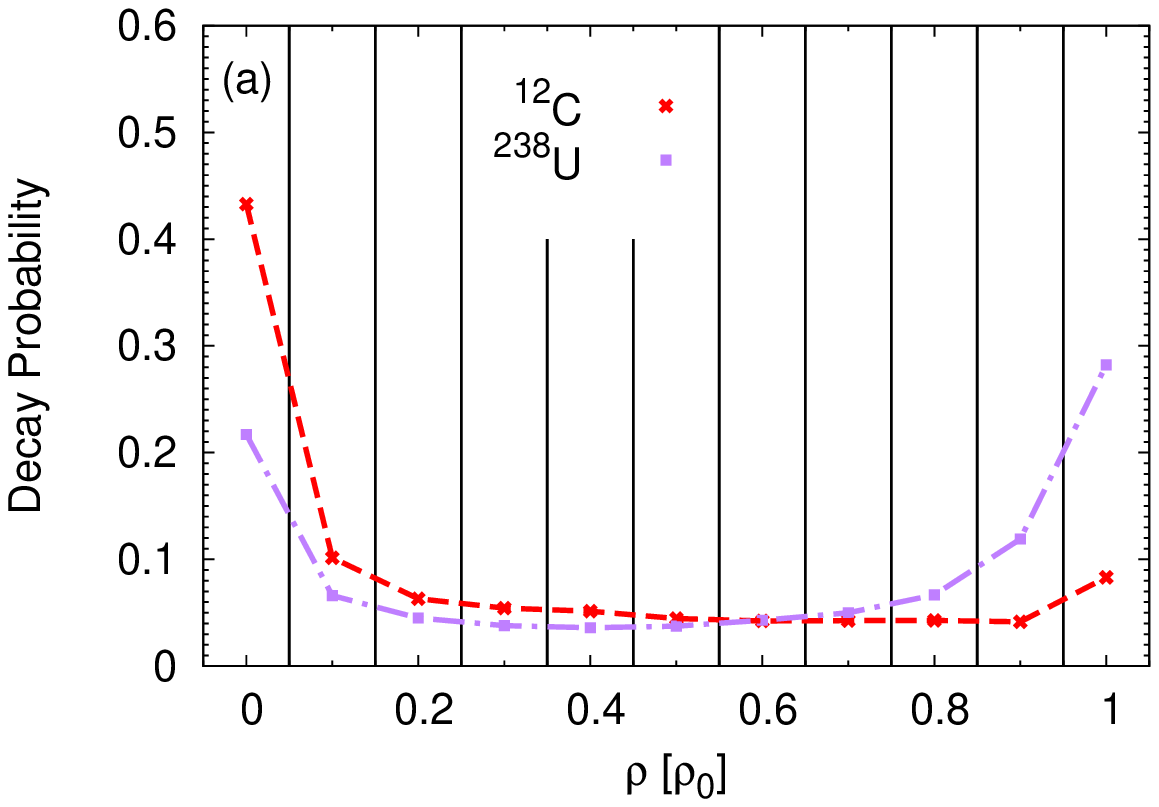}
\includegraphics[scale=0.7]{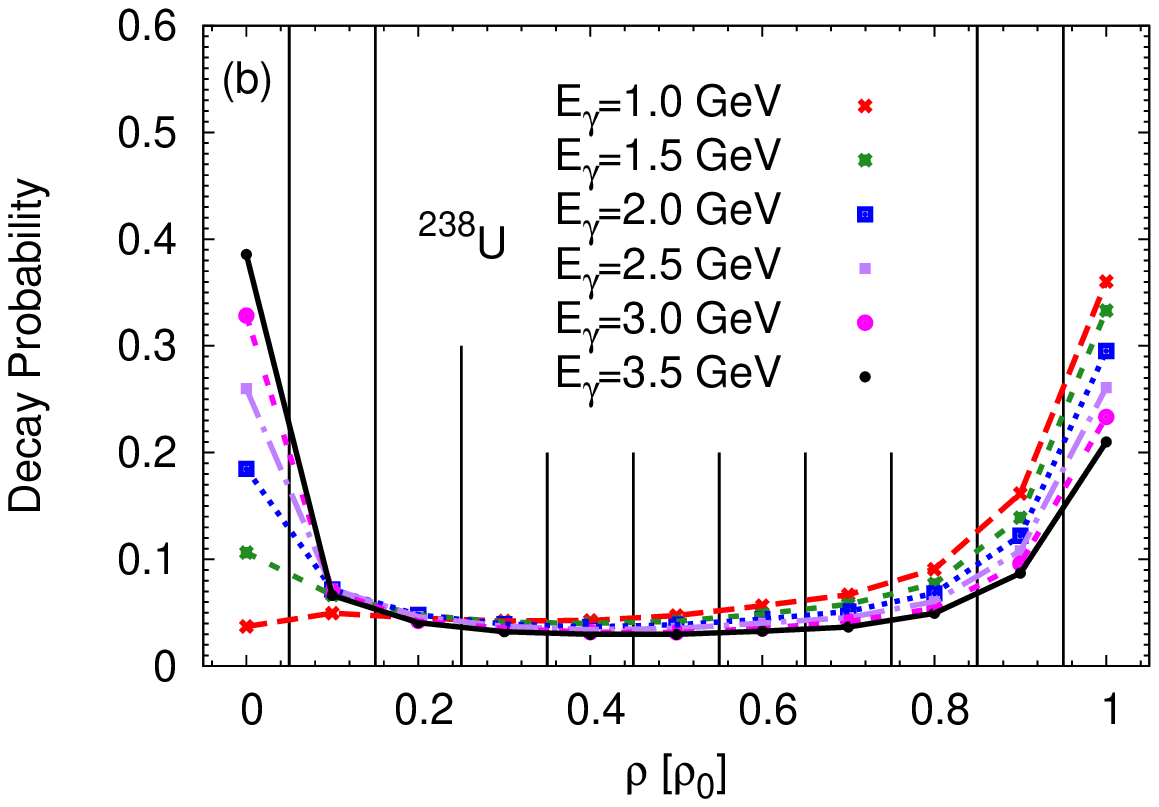}
\caption{(Color online) Probability distributions (normalized to one) of 
the nuclear density at the $\rho$-meson decay points, 
$dP_{\rm decay}/ d\varrho_N$; 
left panel: using an input photon-energy spectrum as in CLAS for two 
differrent nuclei; right panel: using a $^{238}\text{U}$ target for
different photon energies. Vertical lines indicate the bins used for 
the histogram.}
\label{fig_density}
\end{figure*}
We first recapitulate the main components of our model as constructed
in Ref.~\cite{Riek:2008ct}. The elementary photoproduction amplitude, for
the process $\gamma\,p\,\rightarrow\,e^+\,e^-\,p$, is based on the
meson-exchange model developed by two of us~\cite{Oh:2003aw}, which
accounts for $\sigma$, $f_2$, $\pi$, and Pomeron $t$-channel exchange as
well as nucleon $s$- and $u$-channel pole diagrams. This model gives a very
good description of the total $\rho$ production cross section for incoming
photon energies $E_\gamma\ge 2.5$~GeV. Since the input spectrum used by the
CLAS Collaboration at JLab contains photon energies down to $\sim$1\,GeV,
we extended the model by including $s$-channel baryon-resonance
excitations. By implementing the same set of resonances (with the same
coupling constants and form factors) that were used in the calculation
of in-medium $\rho$ spectral function~\cite{Rapp:1997ei,Rapp:1999us},
a good description of the low-energy part of the $\rho$-production cross
section was found. The description of the angular distributions (which
are forward dominated for $t$-channel exchanges) also improves because
the resonance decays generate significant strength at large scattering
angles. This provides a realistic starting point for the
production mechanism on nuclear targets and establishes consistency between
the production amplitude and the medium effects induced by the subsequent
propagation of the $\rho$ meson through the nucleus.
In addition to direct resonance excitations, $\rho+N\to B^*$, the model for
the in-medium $\rho$ propagator includes the dressing of its 2-pion cloud
via nucleon-hole ($NN^{-1}$) and delta-hole ($\Delta N^{-1}$) excitations.
When translated into contributions to the $\gamma N\to\rho N$ cross sections,
the pion cloud dressing corresponds to $t$-channel pion exchanges (with
either nucleon or delta final states), as well as contact terms dictated
by gauge invariance. The $\sigma$, $f_2$, and Pomeron exchanges are not
included in the in-medium $\rho$ spectral function, where they would most
notably contribute at large 3-momentum of the $\rho$ meson. Since
the outgoing $\rho$ momenta
in the CLAS kinematics are typically in a range around $q \simeq 1.5$--$2$~GeV,
it is worthwhile to estimate by how much the broadening in the $\rho$ spectral
function may be underestimated at these momenta. To do this, we neglect nuclear
Fermi motion and employ the schematic expression relating the total absorption
cross section to the imaginary part of the in-medium $\rho$ selfenergy,
\begin{equation}
\Gamma_\rho =- {\rm Im} \Sigma_\rho /m_\rho
            = v_{\rm rel} \ \varrho_N \ \sigma_{\rho N}^{\rm tot} .
\end{equation}
Assuming a (non-resonant) $\rho N$ cross section of 40~mb for a $q=2$~GeV
$\rho$ imparting on a nucleon at rest, $v_{\rm rel}=0.9$, and using
$\varrho_N=0.5\varrho_0^{}$, we find $\Gamma_\rho \simeq 60$~MeV. This is
to be compared to the in-medium $\rho$-meson width obtained for the model
employed here, amounting to $\sim$40\,MeV (for $q=2$~GeV,
$\varrho_N=0.5\varrho_0^{}$, see, e.g., Fig.~4 in our previous
article~\cite{Riek:2008ct}). This estimate suggests that we may underestimate
the $\rho$ width at high momenta by up to 20~MeV.\footnote{However, smaller
momenta contribute and  the ``average" density probed by the CLAS
experiment is
$\sim$$0.4\varrho_0$ for $^{56}$Fe targets; see also later discussion.. }

With reasonable control over the elementary $\rho$-production amplitude
on the nucleon, we applied the above-constructed model to dilepton
invariant-mass spectra off nuclei as measured by the CLAS Collaboration.
The basic assumption in this approach is that the dilepton production
process can be ``factorized" into a 2-step process, namely, $\rho$ production
on a single nucleon followed by its in-medium propagation and final decay.
The pertinent mass-differential cross section for a fixed
incoming photon energy is then given by
\begin{eqnarray}
&& \hspace{-0.45cm}\left\langle\frac{d\sigma}{d\,M}\right\rangle_{A}(E_\gamma,M)
=\frac{m_N^2\,M}{(2\pi)^2\,\varrho_{A}^{}} 
\nonumber\\
&& \hspace{-0.45cm} \times\,\int\frac{d^3p}{(2\pi)^3}\frac{d^4q}{(2\pi)^4}
\frac{d^{4}p^{\prime\,}}{(2\pi)^4}\frac{e^2\,g^2}{(k^2)^2\,m_\rho^4}
\frac{(-{\rm Im}\,\Sigma_{\gamma\,\rightarrow\,e^+e^-}^{\rm vac}(q))}
{2\,\sqrt{(p\cdot k)^2}}
\nonumber\\
&& \hspace{-0.45cm} \times\,\delta(q^2-M^2)\,\delta(p^{\prime\,2}-m_N^2)\,
\delta^4(k+p-q-p^\prime)\,\Theta\left(k_F-|\vec{p}\,| \right)
\nonumber\\
&& \hspace{-0.45cm} \times\,\Theta\left(|\vec{p}^{\,\prime}|-k_F \right) \,
\sum\limits_{m_s,m_{s^\prime},\lambda}T^\mu(k,p,q)\,
\left(T^\nu(k,p,q)\right)^\dagger\,
\nonumber\\
&& \hspace{-0.45cm} \times\,
\left\{P^L_{\mu\nu}(q)\,|G_\rho^L(q)|^2+P^T_{\mu\nu}(q)\,|G_\rho^T(q)|^2\right\},
\label{sigmaF}
\end{eqnarray}
with $k=(E_\gamma,\vec k)$, $p=(E_p ,\vec p\,)$, $p'=(E_{p'} ,
\vec p^{\,\prime}\,)$,
$q=(q_0 ,\vec q\,)$ being the 4-momenta of the incoming photon, incoming 
nucleon, outgoing nucleon (all on-shell) and outgoing $\rho$ (i.e., 
dilepton), respectively. Furthermore, $T^\mu$ is the $\rho$ photoproduction
amplitude off the nucleon, $G_\rho^{L,T}$ are the longitudinal and
transverse electromagnetic correlators encoding the in-medium $\rho$
propagator,\footnote{For example, $G_\rho = (m_\rho^2/g_\rho) D_\rho$ in
the vector dominance model (VDM). The present approach accounts for
corrections to VDM in the baryon sector as described in
Ref.~\cite{Rapp:1997ei}.}
and $P^{L,T}$ are the standard 4-dimensional projection operators.
As is usual in many-body theory, the in-medium width (imaginary part of
the selfenergy) of a particle is calculated in terms of the imaginary
part of the {\em forward} scattering amplitude on particles of the
medium, corresponding to absorption in the forward direction. These processes 
include finite-angle elastic scattering. Thus, Eq.~(\ref{sigmaF}) does
not include the dilepton decay of $\rho$-mesons after being scattered 
elastically at finite angle. There are several reasons why this contribution 
can be expected to be small. For example, the nucleon on which the scattering
occurs is either Pauli-blocked (not for forward scattering!) or ejected
from the nucleus (i.e., corresponding to an extra particle in the
final state). Formally, one would have to calculate such a contribution 
by inserting an extra $\rho$-$N$ scattering amplitude into Eq.~(\ref{sigmaF}),
together with an extra folding over a nuclear Fermi distribution of the 
struck nucleon. Thus, this process is formally of higher order in density.
Given the fact that the in-medium part of the $\rho$ width is basically
linear in density (e.g., cf., Fig.~11 in Ref.~\cite{Rapp:2009yu}), higher
orders due to elastic finite-angle scattering should be suppressed as
well. Of course, ultimately, one should compare our predictions for total
dilepton production cross sections to absolutely normalized data, which,
unfortunately, are not available at this point.

With a realistic photon input-spectrum consisting of six bins of photon
energies, $E_\gamma = 1.0$, $1.5$, $2.0$, $2.5$, $3.0$ and $3.5$~GeV with
relative weights of $13.7$, $23.5$, $19.3$, $20.1$, $12.6$, and 10.9\,\%,
respectively, a fairly good description of the excess mass
spectra\footnote{An ``excess" mass spectrum is defined as the signal
spectrum (after combinatorial background subtraction) with the
(narrow) $\omega$ and $\phi$ peaks subtracted. The remaining signal is
then associated with the $\rho$-decay contribution.}
for deuteron and iron targets has been found~\cite{Riek:2008ct}.
For the full in-medium propagator of the $\rho$ encoded in $G_\rho$ the
input density has to be specified. In Ref.~\cite{Riek:2008ct} this was
 done rather schematically by evaluating an average density at the
$\rho$'s final decay point, implying that the $\rho$-meson propagator
quickly relaxes to the surrounding medium.
The (average) density at the decay point was estimated assuming
(i) a Woods-Saxon profile for the nucleus,
(ii) a uniform sampling of the nucleus for the production point
(i.e., the incoming photon reacts equally likely with each nucleon),
and (iii) a $\rho$-meson path length, collinear with the incoming photon
momentum and estimated using an average $\rho$ momentum of $q\simeq2$~GeV
(determining its velocity), and an average in-medium $\rho$ lifetime of
$\bar{\tau}_\rho=1/\bar\Gamma^{\rm med}=1/200$~MeV~=1\,fm/$c$. The resulting
average density for $^{56}$Fe turned out to be $\sim$0.5$\varrho_0$,
but the variations in the normalized spectra are small when varying this
density within $\pm 0.1\varrho_0$ (Note that, at
these densities and 3-momenta, the width of the microscopic spectral
function is indeed around 200\,MeV, which is consistent with the average
lifetime used for the density estimate.).

Here we improve upon the above procedure in several respects.
First, the width, and thus the decay time, of the $\rho$ meson depends
on its 3-momentum. We replace the average value used before with the
explicit momentum dependence given by our microscopic model for the propagator.
In particular, slow $\rho$ mesons decay faster than fast ones that
consequently travel farther. This could be significant in view
of the rather large range of photon energies (between $0.6$~GeV and
$3.8$~GeV) as used by the CLAS Collaboration~\cite{CLAS-priv}.
Second, the assumption that all $\rho$ mesons of a given velocity decay
after a fixed (average) travel distance, $L$, is subject to corrections.
Here, we include the exponential decay characteristics as
\begin{eqnarray}
\frac{dN}{dt}=\exp\left(-\Gamma[\varrho_N(\vec{r}\,),v]\,t\right) \ ,
\end{eqnarray}
implying that even for identical kinematics and production point the
path length varies and thus different densities at the decay point are
probed.
The initial creation points of the $\rho$ meson are still distributed
according to a realistic density profile for each
nucleus~\cite{DeJager:1987qc} (weighted by volume).
Following this procedure we obtain the decay points and thus a
distribution determining how many $\rho$ mesons decay at a given density.
The final spectrum then follows using this distribution in the electromagnetic
correlator (rather than an average density).

\begin{figure*}[!t]
\includegraphics[scale=0.7]{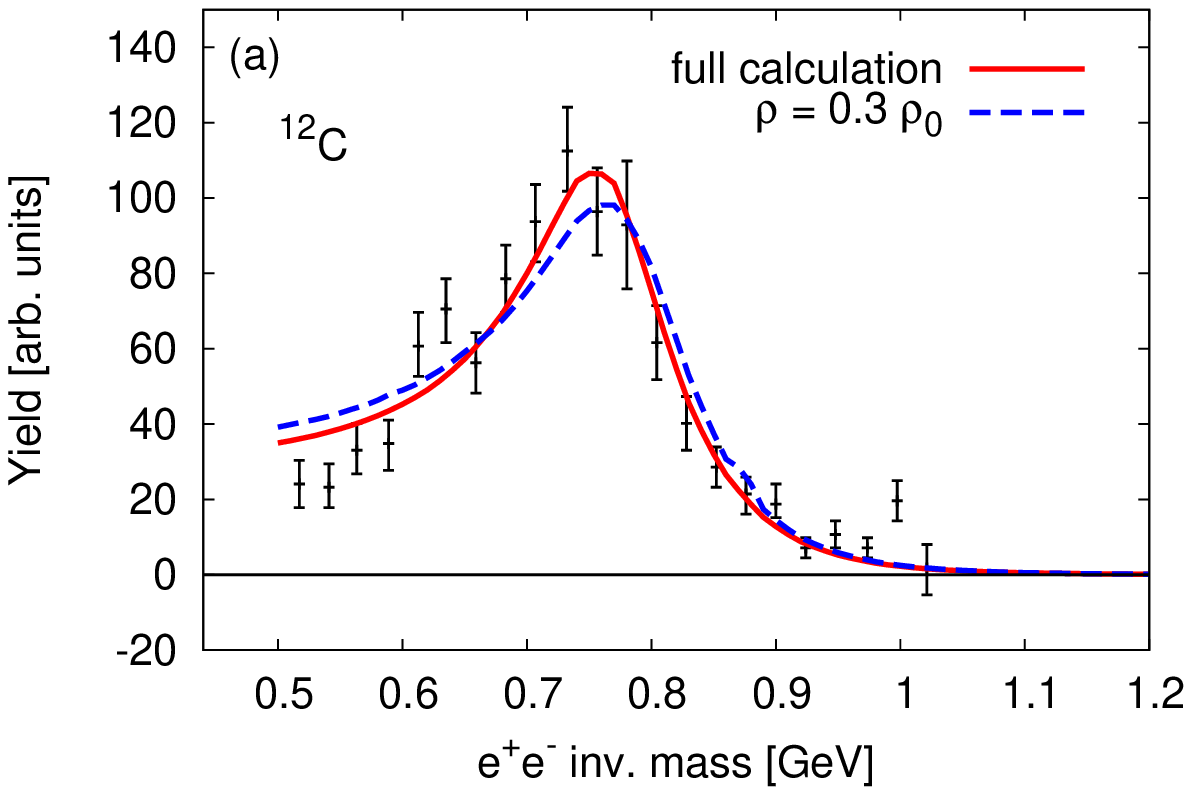}
\includegraphics[scale=0.7]{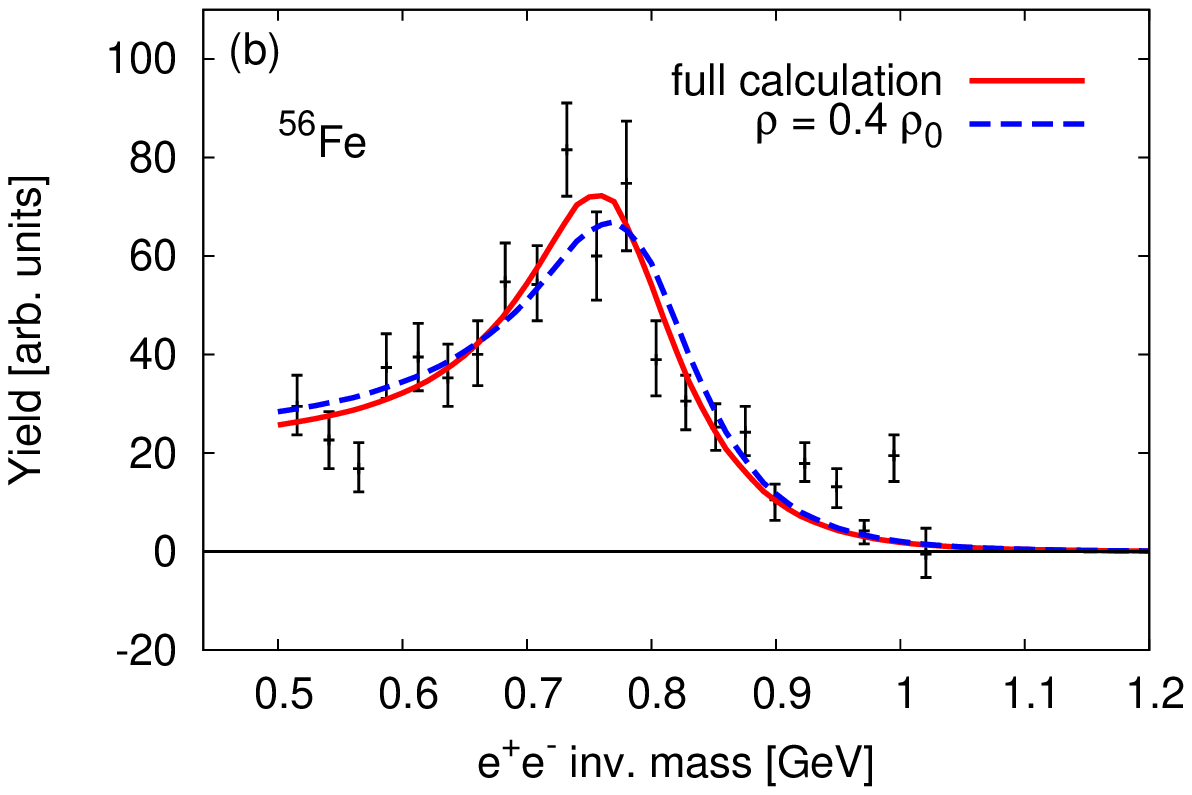}
\includegraphics[scale=0.7]{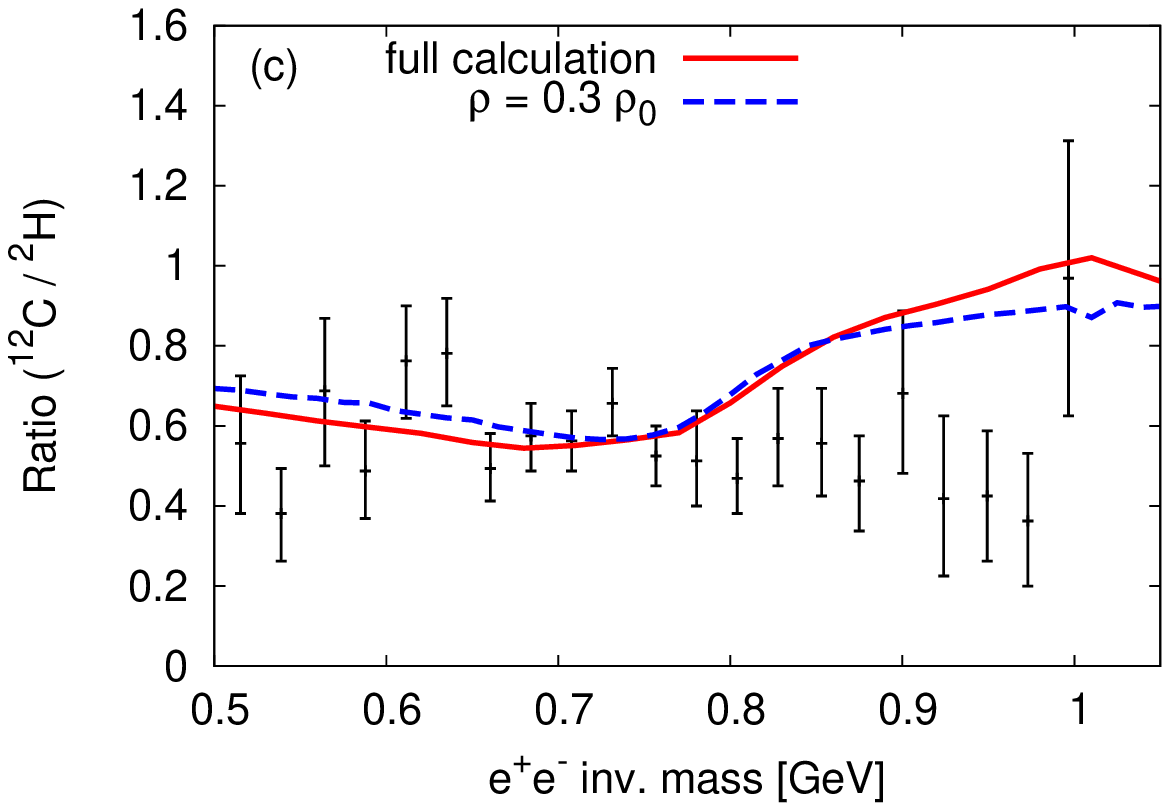}
\includegraphics[scale=0.7]{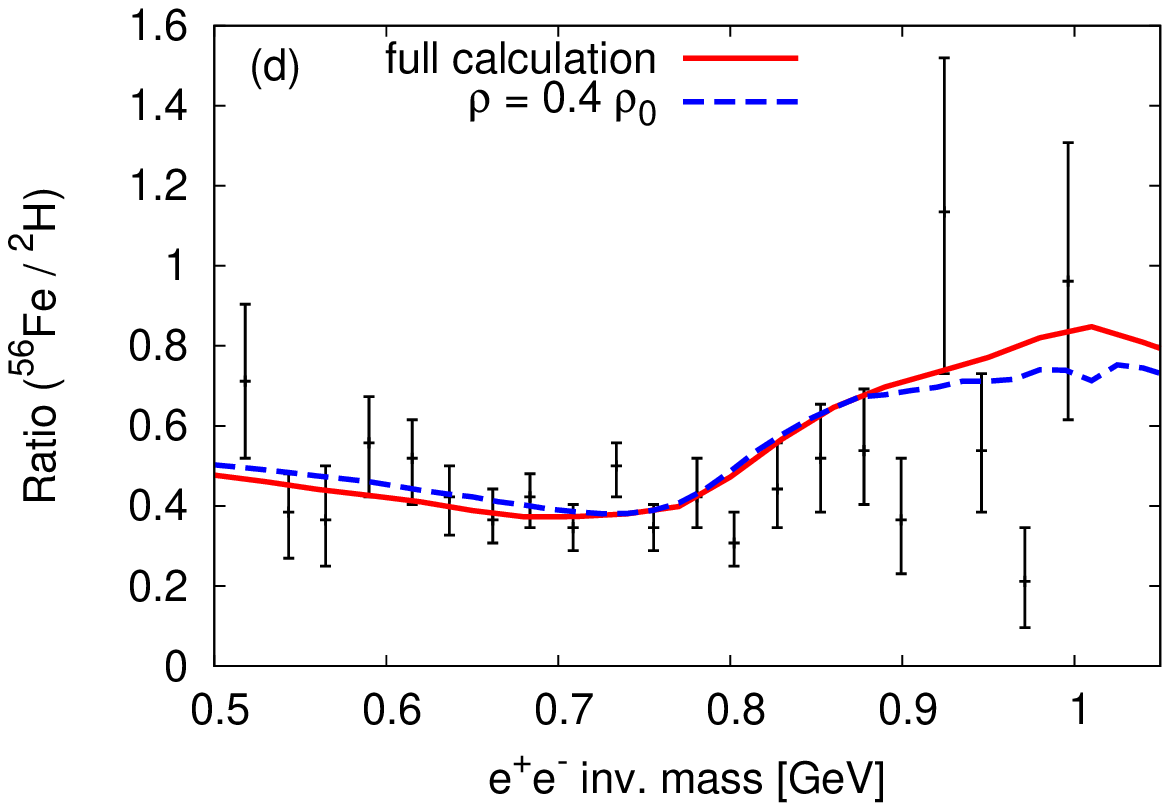}
\caption{(Color online) Excess mass spectra for $^{12}$C and $^{56}$Fe
(upper left and right panel, respectively) and ratios of these spectra
to the deuterium spectrum (lower panels; in the ratios, both numerator 
and denominator are normalized to the corresponding target-mass number, 
$A$). The full calculation using explicit density distributions (solid 
lines) is compared to simplified calculations using a single average 
density.}
\label{fig_M-spec}
\end{figure*}

In Fig.~\ref{fig_density} we show the resulting distributions of
$\rho$-decay points as a function of nuclear density, for a small and 
a large nucleus integrated over the initial photon energy spectrum as 
specified above (left panel), and for a heavy nucleus resolved into 
different initial photon energies (right panel). In both distributions 
we observe a clustering of the decay points at the low and high ends 
of the available densities. In particular, the absolute maximum shifts 
from the lowest density bin ($\varrho_N^{} \simeq 0$) for small nuclei 
to the highest one ($\varrho_N^{}\simeq \varrho_0^{}$) in heavy nuclei, 
and similarly when going from high to low photon energies.
The two-hump structure of the distributions reflects the rather sharp
transition in the nuclear density profiles from a dense interior to the 
vacuum, with a relatively thin surface layer. For the CLAS input photon 
spectrum (which ultimately determines the $\rho$'s velocity and the 
time dilation of its decay time), the use of a heavy nucleus ($^{238}$U)
compared to a light one ($^{12}$C) makes a big difference, by increasing 
the interior decay fraction by a factor of $\sim$3 while decreasing the 
outside decays by a factor of $\sim$2 (see left panel of 
Fig.~\ref{fig_density}).  Reducing the photon energy is particularly 
efficient in suppressing outside decays (see right panel of 
Fig.~\ref{fig_density}), while the exponential decay characteristics 
induces a good fraction of interior decays for relatively fast $\rho$ 
mesons.  The asymmetries in the densities of the decay points are the 
main difference compared to the use of an average density in our earlier 
work~\cite{Riek:2008ct}, which is at the origin of the slight variation 
in the results shown below.

\section{Signatures of Medium Effects
\label{sec_signatures}}
In Sec. III A we first discuss the updated inclusive dilepton mass spectra in
comparison to the CLAS data, as following from the improved
evaluation of decay-point densities as described in the previous section.
In the subsequent three subsections we then scrutinize the origin
of the medium effects to identify observables that can
improve our understanding of the medium effects and test consistency
of spectral shapes and absolute yields.

\subsection{Update of Invariant-Mass Spectra and Comparison to CLAS Data}
\label{ssec_update}
In Fig.~\ref{fig_M-spec} we compare our earlier calculations for the
$^{56}$Fe target with our improved calculation and with the CLAS data at
JLab~\cite{Clas:2007mga,Wood:2008ee}. It also includes the comparison
for the $^{12}$C target. The average densities of $0.3 \varrho_0$
for carbon and $0.4 \varrho_0^{}$ for iron correspond to the averages obtained
from the density distributions in the improved scheme (as we will see in
Sec.~\ref{ssec_transp} below, such averages are also estimated from the
transparency ratios for total cross sections). One finds that
the inclusion of the explicit density distribution leads to a small
downward shift of the $\rho$-resonance peak in the spectrum, which
improves the description of the data compared to the calculation
using an average density. The reason for this effect is that $\rho$ mesons
with large momentum,
which predominantly figure into the high-mass part of the spectrum,
now effectively decay at lower densities, implying reduced medium effects,
that is, a  narrower distribution (and a slightly smaller upward mass
shift). This suppresses the high-mass tail of the dilepton spectrum and
increases the peak height.
Overall, the full model gives a good description of the data for both
nuclei and provides a reasonable baseline for an advanced analysis of
in-medium effects. We note that our findings are in good agreement with
the results of the Boltzmann-Uehling-Uhlenbeck (BUU) transport simulations
by the Giessen group~\cite{Effenberger:1999ay,Muhlich:2002tu,Wood:2008ee}
based on the spectral function of Ref.~\cite{Post:2000qi}. Also in this
calculation, a broadening of the $\rho$-meson spectral function by
$\sim$70\,MeV was found without significant mass shift.

\begin{figure*}[t]
\includegraphics[scale=0.7]{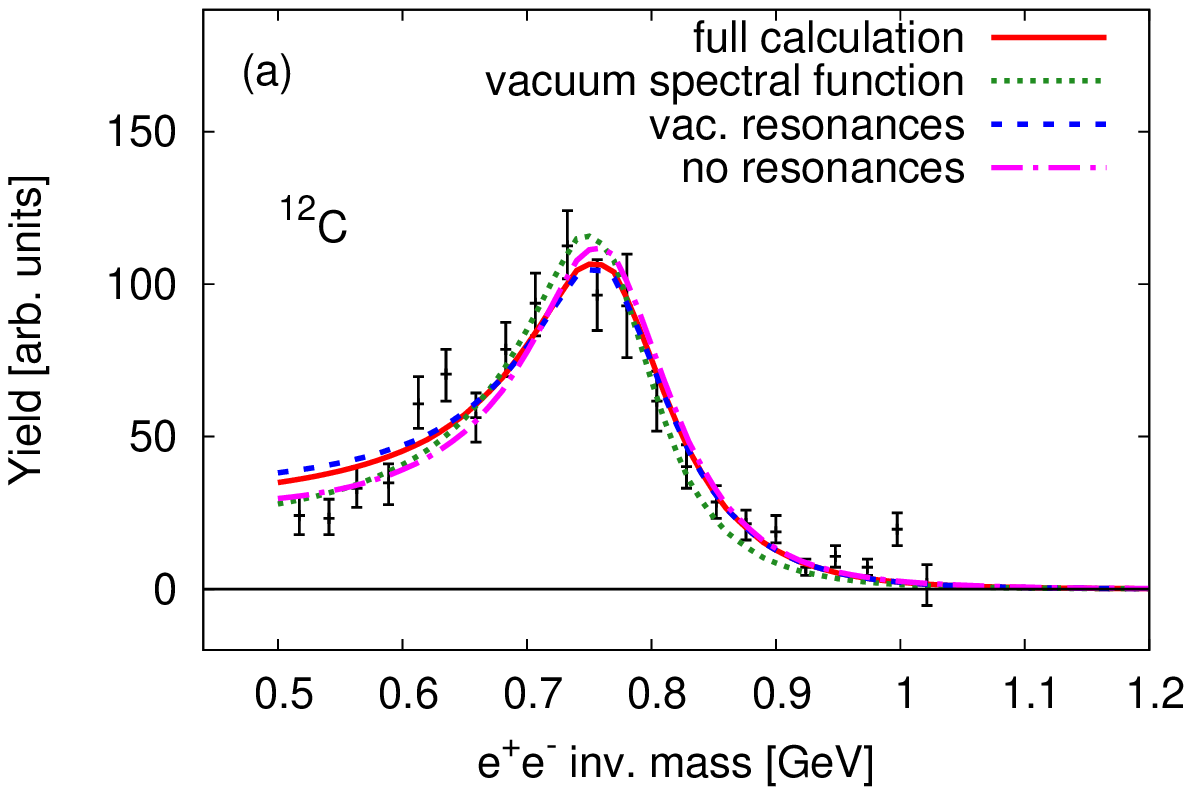}
\includegraphics[scale=0.7]{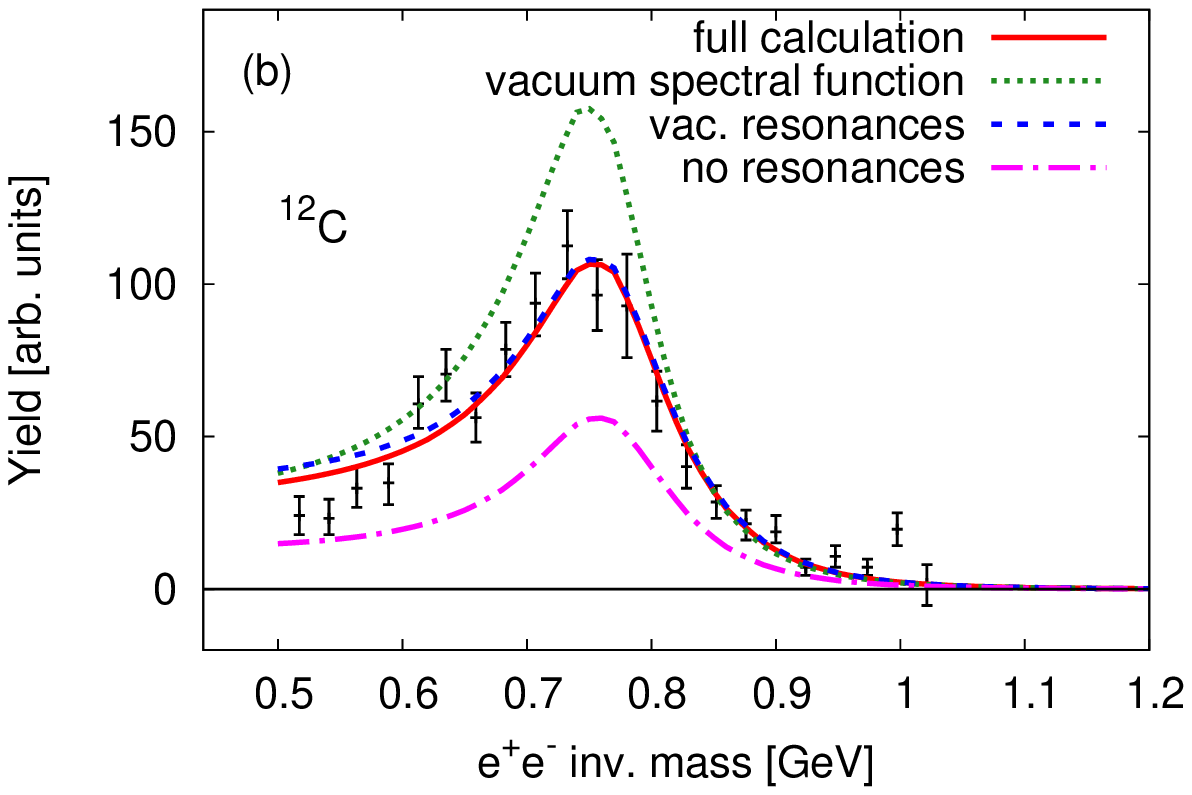}\\
\includegraphics[scale=0.7]{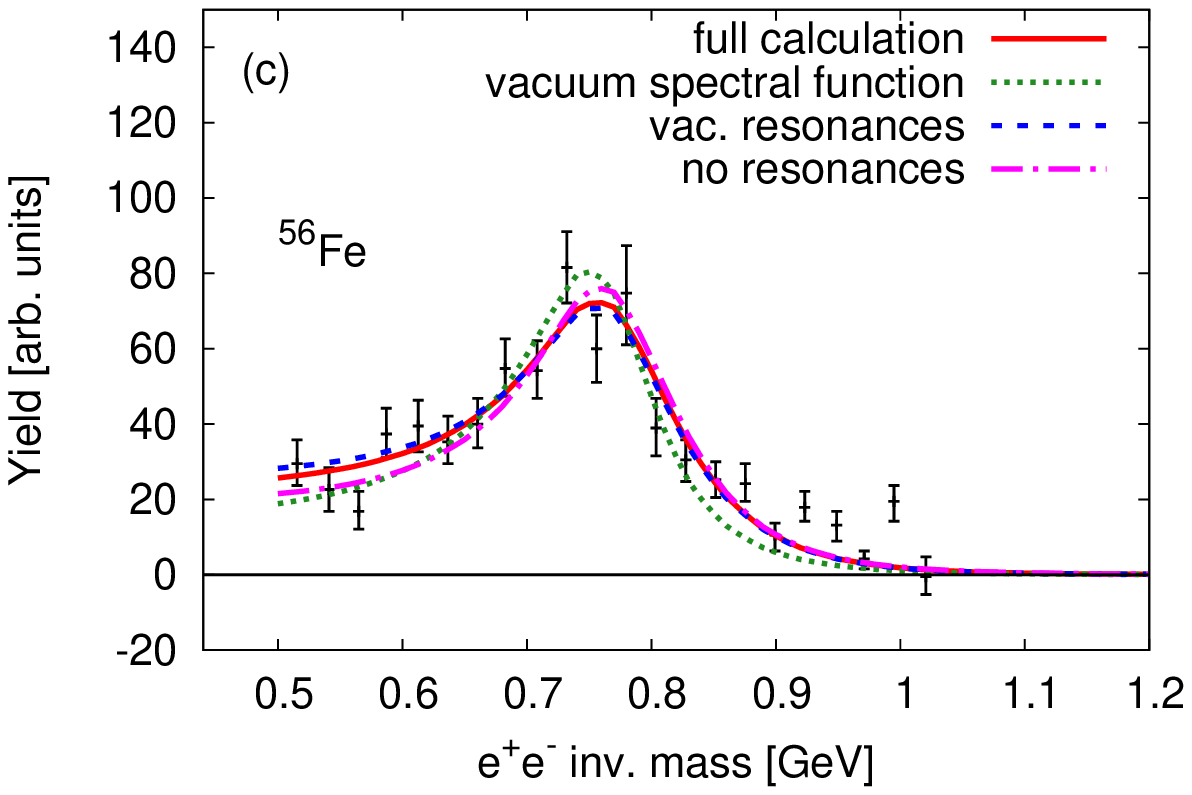}
\includegraphics[scale=0.7]{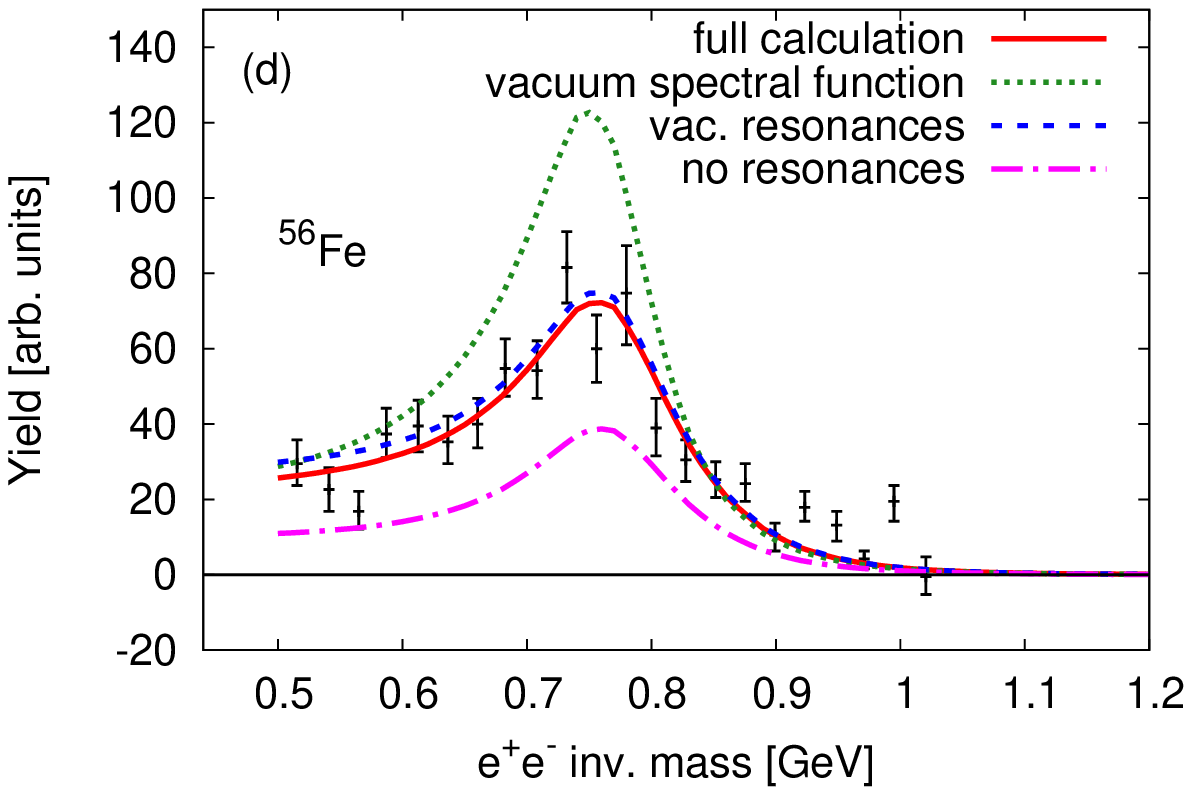}
\caption{(Color online) Calculations for excess mass spectra off
$^{12}$C (upper panels) and $^{56}$Fe (lower panels) targets compared
to the CLAS data~\cite{Clas:2007mga,Wood:2008ee} at JLab. The full
calculations (solid lines) are decomposed by modifying different theoretical
ingredients, by either (i) switching off the baryon-resonance
contributions in the elementary production amplitude (dash-dotted lines)
or (ii) switching off the in-medium widths of the baryon-resonance
contributions in the elementary production amplitude (dashed lines), or
(iii) replacing the in-medium electromagnetic correlator,
$G_\rho$, with the vacuum one (dotted lines). In the left panels all spectra are
($\chi^2$-) normalized to the data, while in the right panels we keep
the normalization of the full model also for the other scenarios.}
\label{fig_M-decomp}
\end{figure*}
Next, we address the influence of different model components on the shape
and magnitude of the invariant-mass spectra. In Fig.~\ref{fig_M-decomp},
we display the results of the calculations in which
(i) the resonances in the production amplitude are switched off
(dash-dotted lines),
(ii) the resonances in the production amplitude are kept as in the vacuum
(i.e., without in-medium broadening as specified in
Refs.~\cite{Rapp:1997ei,Rapp:1999ej}; dashed lines), and
(iii) the vacuum propagator for the $\rho$-meson is employed (dotted lines).
In the left panels of Fig.~\ref{fig_M-decomp}, the integrated yield for each
scenario is normalized to the experimental data (which to our knowledge
are not absolutely normalized) using a $\chi^2$ fit.\footnote{Alternatively,
one could normalize to the integrated strength of the spectrum. The
differences in both procedures are negligible.} In the right panels of
Fig.~\ref{fig_M-decomp}, only the full calculations (solid lines) are
($\chi^2$-) normalized to the data and the same normalization factor is
then applied to all other curves, which maintains the {\em relative}
normalization of all theory curves. The following observations are
made. The impact of the in-medium broadening of the baryon resonances
on the production amplitude is essentially negligible, in both the
shape (left panels) and absolute magnitude (right panels) of the final
dilepton spectra.
Switching off the resonance contributions in the production amplitude
altogether (dash-dotted lines) leads to a slight narrowing of the spectral
shape (left panels) but, more significantly, reduces the absolute yield of
the cross section by more than 50\% (see dash-dotted lines in the right
panels of Fig.~\ref{fig_M-decomp}). This can be easily understood since
the resonances play a significant role in the total $\rho$ photoproduction
cross section on the nucleon for photon energies below
$E_\gamma\simeq 2$~GeV~\cite{Riek:2008ct}.
An important question concerns the sensitivity
of the spectra to the medium effects in the $\rho$-meson propagator.
We investigate this question by replacing the in-medium propagator with the
vacuum one (resulting in the dotted lines in Fig.~\ref{fig_M-decomp}).
The spectral shape narrows, but not as much as one may have expected.
In fact, when normalizing the theory curves to the experimental
data, the resulting spectral shapes for vacuum and in-medium
correlators are not much different for both carbon and iron targets, see
left panels of Fig.~\ref{fig_M-decomp}.
More quantitatively, for the carbon target, the $\chi^2$ per data point
is close to 1 for both in-medium and vacuum $\rho$ propagator; that is, no
significant distinction can be made. For the iron target, we actually have
a preference of $1.08$ over $1.55$ in the $\chi^2$ when using the in-medium
propagator. This is in line with the analyses carried out in
Ref.~\cite{Wood:2008ee} based on BUU
calculations~\cite{Effenberger:1999ay,Muhlich:2002tu} with the
in-medium broadened spectral function of Ref.~\cite{Post:2000qi}, or
on direct fits to the spectra with a Breit-Wigner ansatz for the spectral
function~\cite{Wood:2008ee}.
We recall that the absolute yields of the spectra markedly
increase for the vacuum spectral function compared to the in-medium one 
(which is more systematically quantified in Sec.~\ref{ssec_transp}).

The question we want to address in the following is how we can
better discriminate medium effects from the experimental data.
For example, for the many-body spectral function employed in our
approach the broadening is much more pronounced at low 3-momenta
(relative to the nuclear rest frame). The moderate broadening observed
in the CLAS data is accounted for by the reduced medium effects
at large 3-momentum (following from the relatively soft hadronic
form factors
as determined in Refs.~\cite{Rapp:1997ei,Rapp:1999ej}), as well as
the limited fraction of in-medium decays even for the iron target.
We investigate several ways to better pin down medium effects.
First, one could use heavier nuclei. This would allow more $\rho$ mesons
to decay inside the nucleus and thus augment the medium effects
(recall the left panel in Fig.~\ref{fig_density}). In addition,
the density dependence of the spectral function could be studied.
Second, one could concentrate on lepton pairs with low 3-momentum
which should enhance the decay fraction inside the nucleus because
of the smaller distance they travel from production to decay
point\footnote{This poses formidable experimental challenges since in the
present CLAS set-up only a detection of pairs with momenta above
$\sim$1\,GeV is feasible~\cite{Leupold:2009kz}.}.
A breakdown of the data in several momentum bins would illuminate
the 3-momentum dependence of the spectral function, a critical property
as indicated above.
Third, as the presumably most straightforward option on the experimental
side, one could utilize the absolute magnitude of the total cross section
in terms of the so-called nuclear transparency ratio.

The above-listed three possibilities are elaborated in the following
three subsections \ref{ssec_target}, \ref{ssec_mom}, and
\ref{ssec_transp}, respectively. To facilitate direct comparisons
to (existing or future) CLAS data, we use the same folding over the
incoming photon-energy spectrum as quoted above~\cite{CLAS-priv}, unless
stated otherwise.
Our hope is that several long-standing issues about basic properties
of the $\rho$-spectral function in nuclear matter (see, e.g., the
differences between the spectral functions of
Refs.~\cite{Rapp:1999us,Post:2000qi,Cabrera:2000dx,Lutz:2001mi}) could be
resolved by detailed studies of accurate $e^+ e^-$ spectra
in reactions off nuclei. One of our main goals is to quantify the
notion of ``accurate".

\subsection{\boldmath $A$ Dependence of the Spectral Shape}
\label{ssec_target}
%
\begin{figure*}[t]
\includegraphics[scale=0.7]{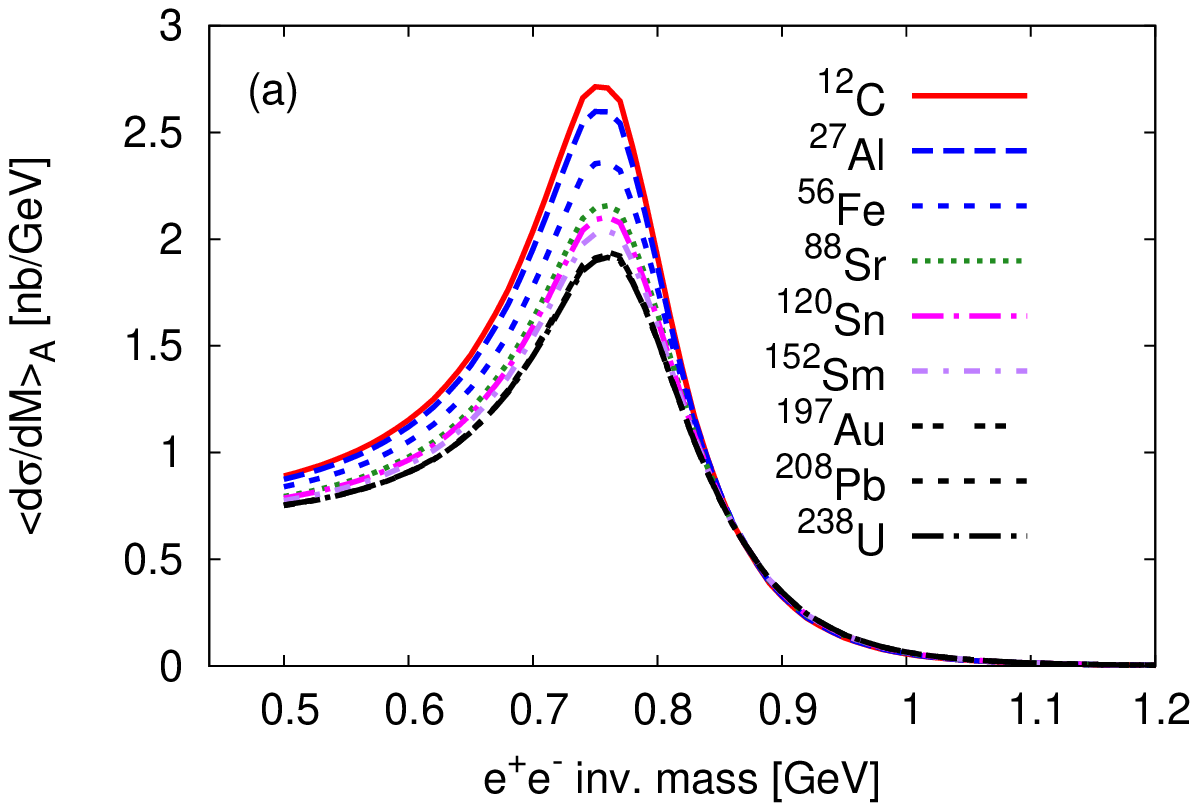}
\includegraphics[scale=0.7]{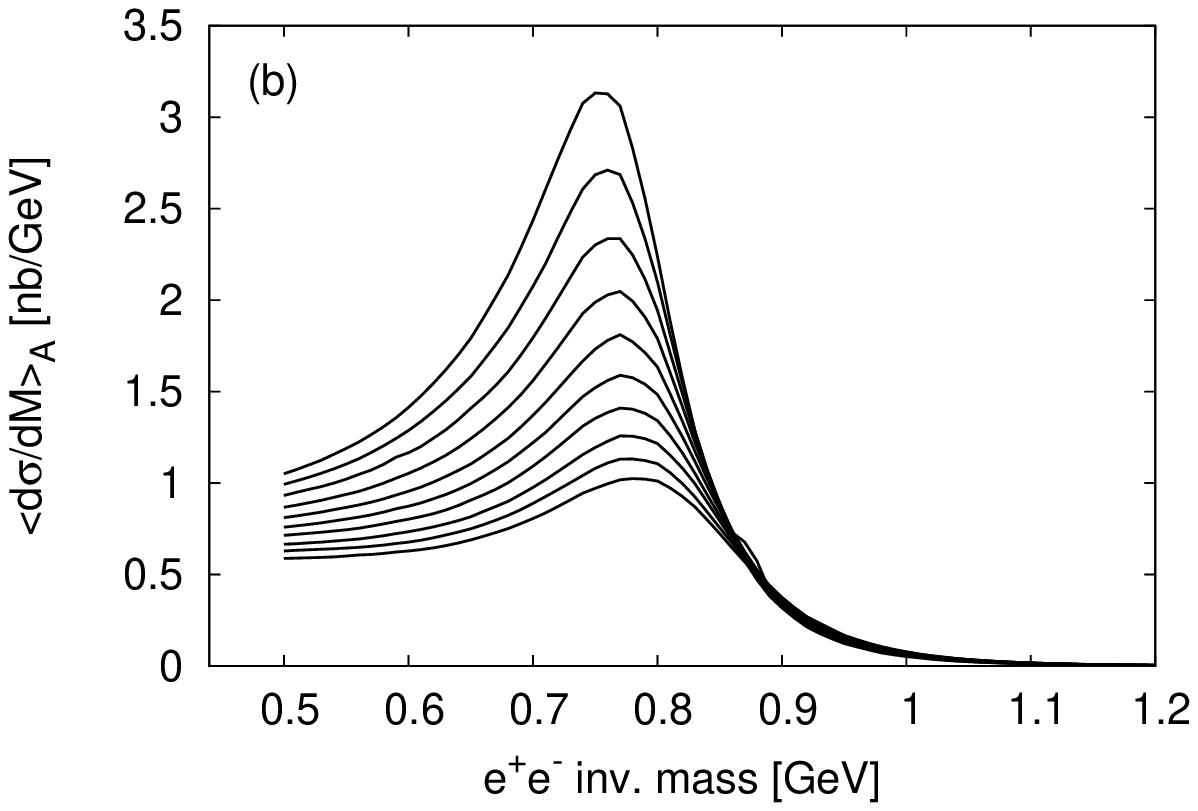}
\caption{(Color online) Absolutely normalized cross sections for excess
mass spectra for several nuclei (left panel) from carbon (uppermost curve)
to uranium (lowest curve) and for several fixed nuclear densities (right
panel) from $0.1 \varrho_0^{}$ (highest curve) to $\varrho_0^{}$
(lowest curve).}
\label{fig_M-Adep}
\end{figure*}
A systematic study of the $A$ dependence of the absolutely normalized
dilepton mass spectra under the CLAS conditions is compiled in the left panel
of Fig.~\ref{fig_M-Adep}. One clearly recognizes
the enhanced sensitivity to the medium effects with increasing $A$ of the
nucleus, mostly due to the simple fact that a larger nuclear volume
increases the fraction of in-medium decays (recall also
Fig.~\ref{fig_density}). The ``gain" in medium effects is quite appreciable
when going from the currently available iron target to Au, Pb or U,
which is more pronounced than the difference between C and Fe.
It instructive to compare the $A$-dependence of the spectra to the
density dependence in infinite nuclear matter,  as shown in the right
panel of Fig.~\ref{fig_M-Adep}. This comparison suggests that for
$A \simeq 200$ targets the inclusive spectral shape (or net broadening)
corresponds to rather moderate {\em average} densities,
$\bar\varrho_N^{} \simeq0.5\varrho_0^{}$.
\begin{figure*}[t]
\includegraphics[scale=0.7]{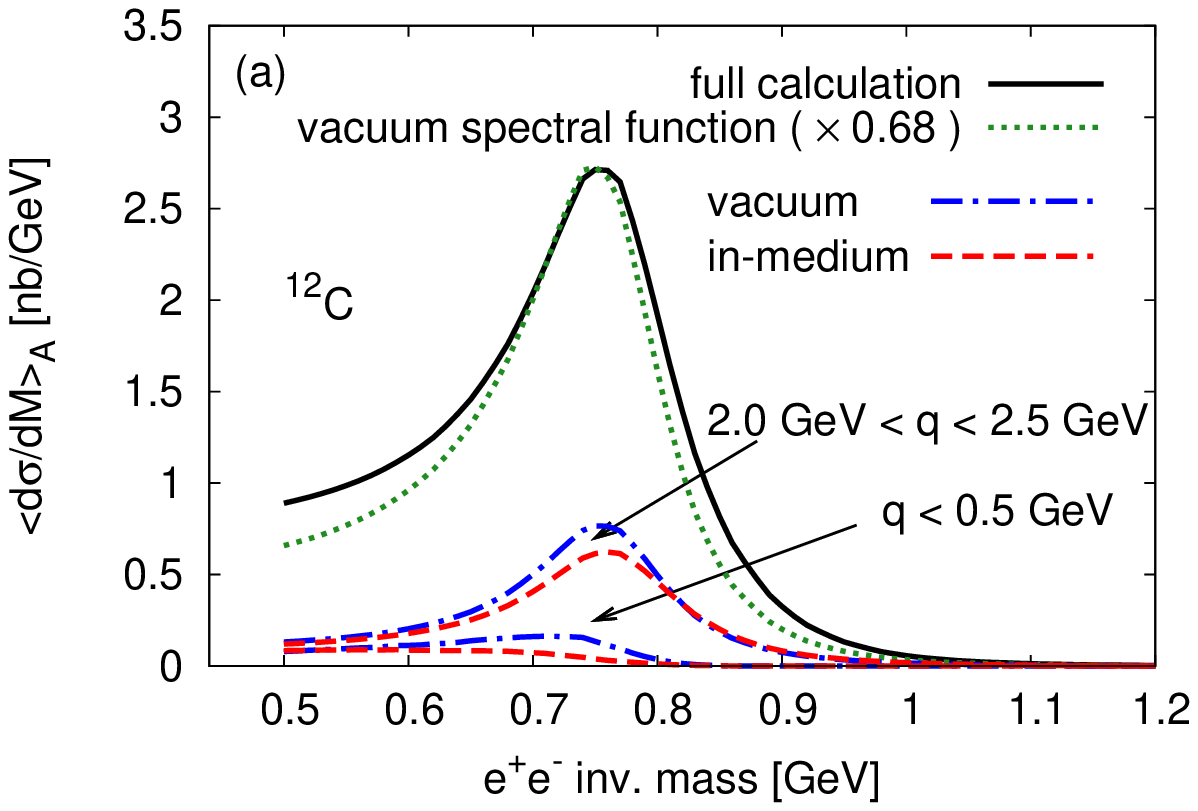}
\includegraphics[scale=0.7]{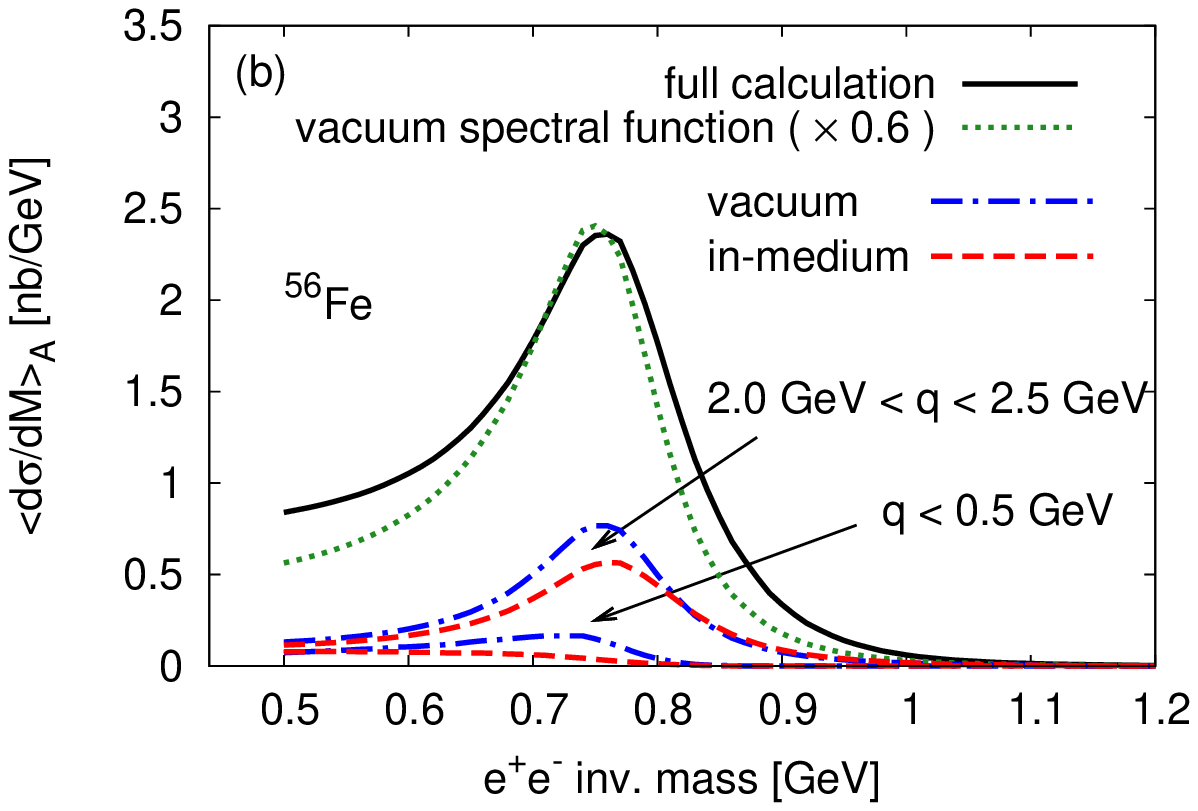}\\
\includegraphics[scale=0.7]{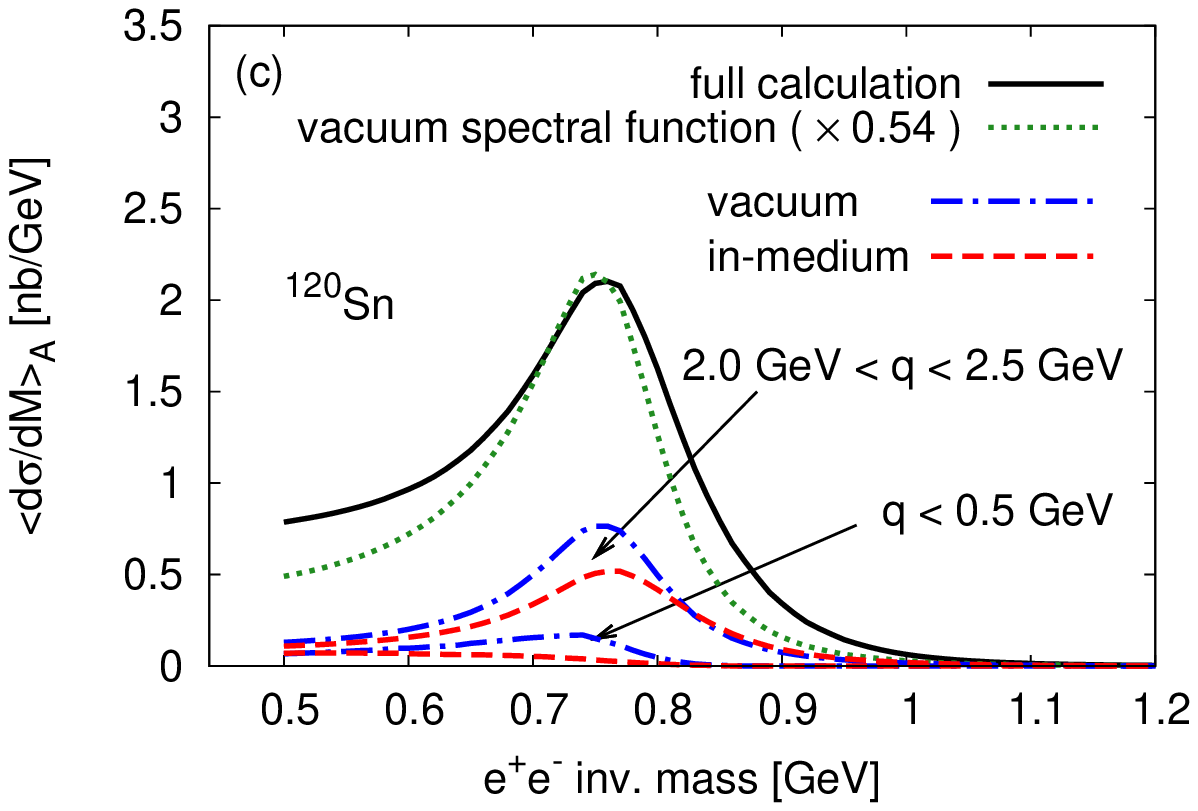}
\includegraphics[scale=0.7]{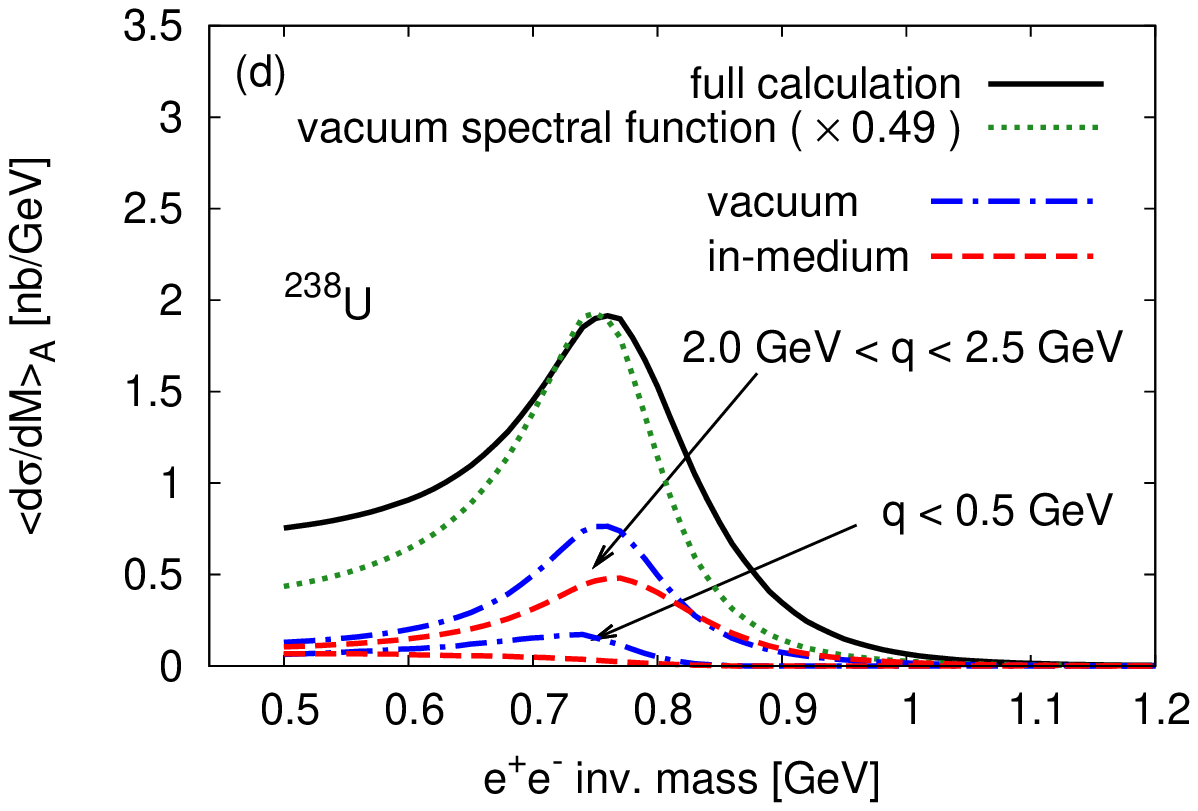}
\caption{(Color online) Excess dilepton invariant-mass spectra for several
nuclei. The full model (solid lines) is compared to
the calculations using the
vacuum $\rho$-propagator (scaled to the peak height of the full calculation
by the indicated factors). Furthermore, absolutely normalized calculations
are shown for in-medium and vacuum spectral functions applying the
indicated momentum cuts to the spectra.}
\label{fig_M-qdep}
\end{figure*}
In an attempt to better quantify the {\em sensitivity} of the
spectral shape to the in-medium broadening, we compare in
Fig.~\ref{fig_M-qdep} the full calculations (solid lines) with calculations
using the vacuum spectral function, where the latter are
normalized such that the peak height coincides with the full calculation
for each of the four nuclei. The sensitivity to medium effects may then
be defined as the enhancement of the full over the vacuum curve
in the low-mass tail, e.g., at $M=0.5$~GeV. The pertinent low-mass
enhancement factors are $1.35$, $1.49$, $1.6$, and $1.73$
for $A=12$, $56$, $120$, and $238$, respectively.


\subsection{3-Momentum Dependence}
\label{ssec_mom}
Next we investigate the option of applying 3-momentum cuts on the outgoing
lepton pair (which in our approximation is equivalent to the 3-momentum
of the $\rho$ relative to the nuclear rest frame). From general
considerations (and our discussions in Sec.~\ref{sec_pro-dec}) we expect
that low-momentum dilepton samples should contain noticeably larger
fractions of $\rho$ decays in the nuclear volume.
Figure~\ref{fig_M-qdep} contains the results where the 3-momentum of the
outgoing $\rho$ is restricted to lie either below $0.5$~GeV or between $2.0$
and $2.5$~GeV. We refer to the former as ``low-momentum cut", while the
latter represents a typical momentum range resulting from the photon
input spectrum used by the CLAS Collaboration (this range
of ``intermediate" momenta accounts for roughly 25\% of the total yield for the full calculation).
The potential benefits of a low-momentum cut are twofold. In addition to
the already mentioned increase of decay probability in the nuclear volume
(which also augments the decays at higher densities), the medium
modifications of our $\rho$ spectral functions increase
appreciably toward smaller momenta. The numerical results displayed
in Fig.~\ref{fig_M-qdep} corroborate the discrimination power of the
low-momentum cut. The shape of the resulting mass spectrum is markedly
different when the full calculation (dashed lines) is compared with a
calculation where the vacuum propagator is used instead (dot-dashed lines),
already for rather light nuclei. The main effect is a quenching of the
$\rho$ peak at its resonance position, while the (absolute values of the)
low-mass cross section turn(s) out to be very similar (the absorption
is approximately compensated by the spectral broadening). Again, one
may try to quantify the sensitivity to the medium effects by defining
an appropriate ratio. The coincidence of the low-mass
cross section suggests the peak ratio as a suitable measure, that is,
the ratio of the resonance peak for the vacuum calculation to the
value at the same mass for the full calculation. For the low (high) momentum cut we find ratios of
$2.4$ ($1.2$), $2.8$ ($1.4$), $4.2$ ($1.5$), and $4.4$ ($1.6$) for $A=12$, $56$, $120$, and $238$,
respectively, which provides a much more increased sensitivity over the
inclusive shape analysis.  Indeed, when comparing the cross sections
for vacuum and in-medium spectral functions
for the above defined ``representative" 3-momentum range, the
sensitivity is substantially reduced.

\subsection{Transparency Ratio (Yield)}
\label{ssec_transp}
%
\begin{figure}[t]
\includegraphics[scale=0.7]{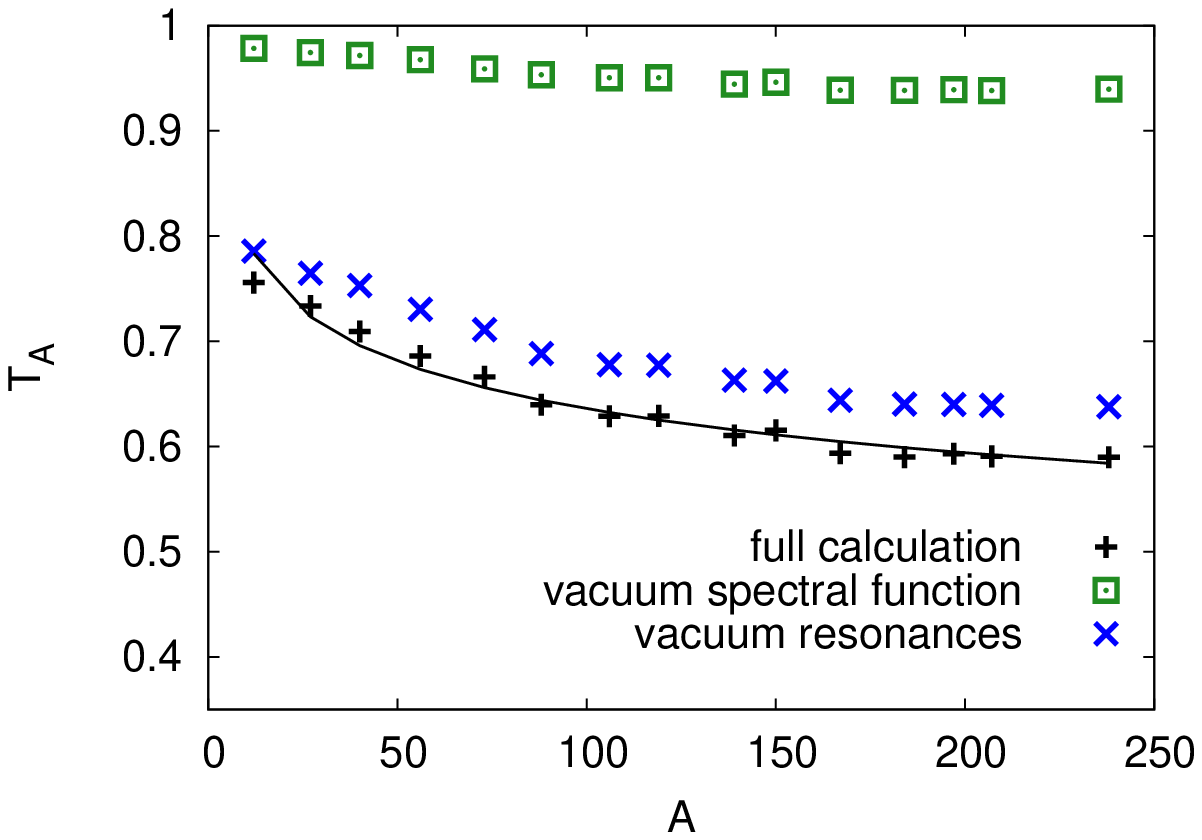}
\caption{(Color online) Nuclear transparency ratio, $T_A$, as a function
of nuclear mass number, $A$. The full calculation is compared to
results when switching off the medium effects on the baryon resonances
in the production amplitude and when using the vacuum $\rho$-propagator
in the electromagnetic correlator, $G_\rho$.}
\label{fig_T-A}
\end{figure}
\begin{figure}[t]
\includegraphics[scale=0.7]{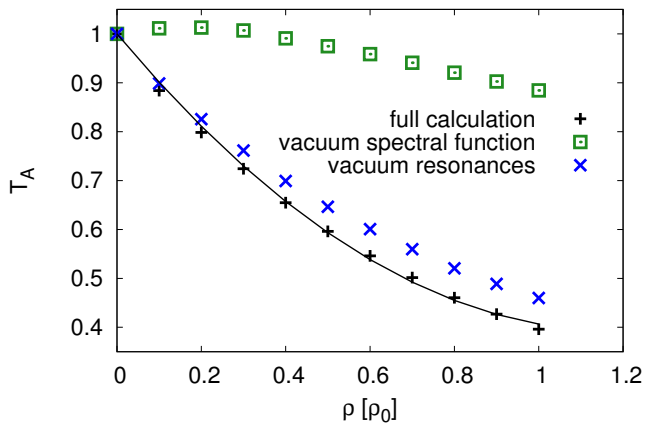}
\caption{(Color online) Same as Fig.~\ref{fig_T-A} but in infinite nuclear matter as a function of density.}
\label{fig_T-rho}
\end{figure}
\begin{figure}[t]
\includegraphics[scale=0.7]{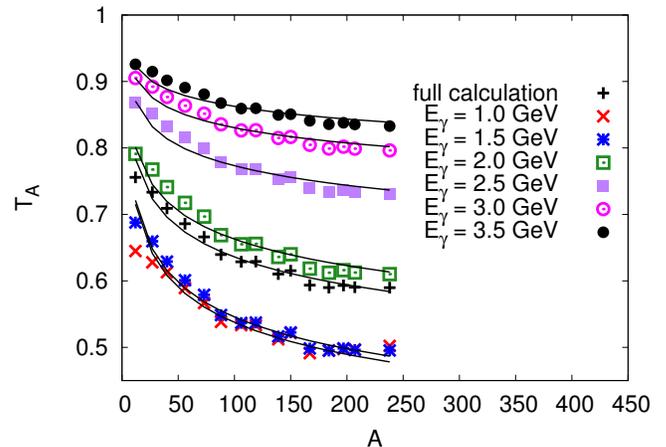}
\caption{(Color online) Transparency ratio as a function of the mass number $A$ for different photon energies.}
\label{fig_T-A-Egamma}
\end{figure}
\begin{figure}[t]
\includegraphics[scale=0.7]{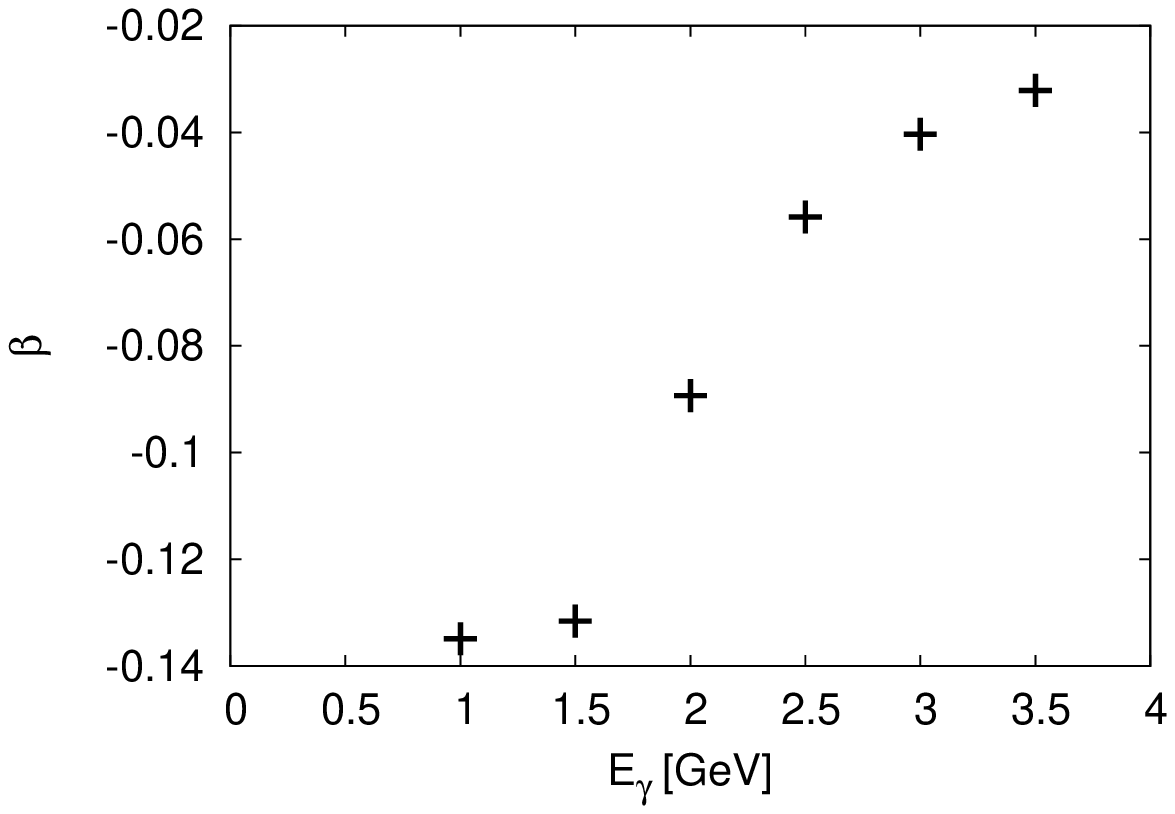}
\caption{(Color online) Fit parameter $\beta$ used to describe the dependence of the transparency ratio on the photon energy.}
\label{fig_para}
\end{figure}
Finally, we turn to our third option to scrutinize medium effects, by
studying the absorption of $\rho$ mesons, rather than the 
spectral shape of the dilepton invariant-mass spectra. The width 
figuring into the mass spectra is of the same physical origin as the
forward absorption of $\rho$ mesons when traveling through the
nuclear medium. Experimentally, the nuclear absorption is usually
quantified by the nuclear transparency ratio,
\begin{eqnarray}
T_A= \frac{\sigma_{\gamma A\,\rightarrow\,\rho X}}
{A\cdot\sigma_{\gamma N\,\rightarrow\,\rho X}}
\simeq
\frac{\int dM \left\langle d\sigma/dM \right\rangle_{\rm{med}}}
{\int dM \langle d\sigma/dM \rangle_{\rm{vac}}} \ ,
\label{T_A}
\end{eqnarray}
which measures the suppression of the nuclear cross section per nucleon,
relative to the cross section on the free nucleon (note that the latter
cross section includes the finite decay width of the $\rho$ in vacuum). 
This quantity was previously measured in nuclear photoproduction of
$\omega$ and $\phi$
mesons~\cite{Ishikawa:2004id,Kotulla:2008xy,Nasseripour:2008zz}, from
which large in-medium widths of $\sim 100$--$150$~MeV have
been extracted (see also
Refs.~\cite{Muhlich:2002tu,Muhlich:2006ps,Kaskulov:2006zc}
for theoretical evaluations). These values are more than an order
of magnitude larger than the respective vacuum widths of $\omega$
($\Gamma_\omega$=8.5\,MeV) and $\phi$ ($\Gamma_\phi$=4.3\,MeV).
This is the main reason why for  narrow vector mesons the transparency
ratio is much more sensitive to medium effects
than invariant-mass spectra (in which the in-medium to vacuum decay
fraction is small). The transparency ratio is less favorable for the
$\rho$ meson where the large vacuum width suppresses the contribution
from vacuum decays (which, in principle, is of course a welcome feature).
In addition, the integration ranges in the transparency ratio,
Eq.~(\ref{T_A}), become more ambiguous for a broad resonance, at least
in practice (i.e., in experiment).
However, we already deduced from Fig.~\ref{fig_M-decomp} that, also for
the $\rho$ meson, the change in the absolute yields may be more pronounced
(i.e., easier to observe) than the medium modifications of the spectral
shape, at least for relatively small nuclei and toward higher momentum.
We therefore computed the nuclear transparency
ratio for the $\rho$ meson within our approach\footnote{Note that our
spectra shown above were already normalized per
nucleon~\cite{Riek:2008ct} so that no extra factor $A$ should be included.}
by integrating the invariant-mass spectra as computed above over
$M=0.5$--$1.1$~GeV.
The results are displayed in Fig.~\ref{fig_T-A} as a function of nuclear
mass number, $A$.
Even for small nuclei a considerable suppression occurs which gradually
becomes stronger for larger nuclei until a ratio of about $0.6$ is reached
for uranium. This is to be compared to the measurements for the $\phi$ and
$\omega$, where the suppression reaches down to $\sim$0.4, even when
normalizing to carbon. Again, the reason for the apparently weaker effect
for the $\rho$ is its large vacuum width of $\sim$140\,MeV, relative to which
changes in the medium are at most a factor of $2$--$3$. On the contrary,
for the $\omega$ and $\phi$ more than a factor of $10$ broadening was extracted and/or
predicted~\cite{Riek:2004kx,Eletsky:2001bb,Wolf:1997ib,Roy:1999mk,Klingl:1997kf,Rapp:2000pe,Kampfer:2003sq}, which leads to much stronger effects in the
transparency ratio.
For completeness, we also investigated different components in
our model calculation. When leaving out the in-medium width increase of
the baryon resonances in the production amplitude, $T_A$ exhibits a small
enhancement of about ~$0.025$--$0.05$ owing to a small increase in the production
cross section. This effect was essentially invisible at the level of the
invariant-mass spectra and underlines the sensitivity of the transparency
ratio to even small variations in the calculation. The main suppression
mechanism is, as expected, caused by the absorption of the propagating
$\rho$ meson, as illustrated by using the vacuum  propagator in which
case $T_A$ is close to one (the small residual suppression being caused
by the medium modifications in the production cross section, essentially
due to the in-medium width of the resonances).

It is instructive to evaluate the nuclear transparency ratio as a function
of nuclear density in infinite nuclear matter\footnote{Note that we apply
the notion of ``infinite matter" only to the strong-interaction parts
of the calculation, i.e., we still pretend that the medium remains
transparent to the electromagnetic probe so that dileptons can
leave without
re-interactions.} (cf.~Fig.~\ref{fig_T-rho}).
This enables yet another
angle at the question of what typical (``average") densities might
be representative for a given nucleus.
For example, we find that to obtain the same $T_A$ for $^{12}$C ($^{56}$Fe)
an infinite density of about ~0.2--0.3$\varrho_0$ (0.3--0.4$\varrho_0$)
should be used, which is consistent with the values obtained with the
method used in Ref.~\cite{Riek:2008ct}.

Of further practical interest is the dependence of $T_A$ on the incoming
photon energy. This information is displayed for our approach in
Fig.~\ref{fig_T-A-Egamma}, which shows that the suppression markedly
increases with decreasing photon energy. Lower photon energies produce
$\rho$ mesons of smaller 3-momentum which in turn increases the time
spent in the nuclear medium and thus provides longer durations of
suppression.
In fact, for incoming photon energies below $E_\gamma\simeq1.5$~GeV,
directed onto $A\simeq200$ targets, $T_A$ has approached the
infinite-matter value for saturation density (0.4) within about ~20\%.
This reiterates once more the power of low-momentum cuts on the lepton
pair. To quantify the $A$ dependence of the transparency ratio in a
functional form, we fitted our results in Figs.~\ref{fig_T-A} and
\ref{fig_T-A-Egamma} with the standard power-law ansatz, $T_A=A^{\beta}$.
For the individual incoming photon energies, the extracted values of
$\beta$ are plotted in Fig.~\ref{fig_para}. For the full calculation
using the weighted input spectrum corresponding to the JLab beam
we find a value of $\beta=-0.0983$. Its magnitude is significantly
smaller than for the available $\omega$ and $\phi$ data, but, surprisingly,
comparable to $J/\psi$ suppression in fixed-target $p$-$A$ collisions
at high energy (see, e.g., the recent compilation in
Ref.~\cite{Rapp:2008tf} where the exponent $\alpha\equiv1+\beta$;
however, this is most likely a coincidence since the physics of $\rho$
and $J/\psi$ nuclear suppression is presumably quite different).
Note that the overall $\beta$-value for the incoming photon
spectrum would correspond to an ``average" photon energy of slightly
below $E_\gamma=2$\,GeV (the weighted mean of our input distribution is
$\bar E_\gamma=2.14$\,GeV).  We finally remark that especially for
light nuclei like carbon a more complicated fit function would be
required to accurately reproduce the calculated values of $T_A$.

\section{Conclusions}
\label{sec_concl}
We have conducted a theoretical analysis of invariant-mass spectra and
total cross sections of dileptons in nuclear photoproduction for incoming
photon energies in the few-GeV regime as recently measured by the CLAS
Collaboration at JLab. Our approach is based on an effective hadronic
model in which we combined in a consistent way an earlier constructed
elementary production amplitude with an in-medium $\rho$ propagator.
The former describes well the $\rho$ production off the nucleon, while
the latter was successfully applied to dilepton spectra measured
in high-energy heavy-ion collisions at the CERN-SPS.

We first improved on our earlier calculations~\cite{Riek:2008ct} by
constructing a more realistic density distribution for the decay points
of the $\rho$ meson which, in particular, accounts for the 3-momentum
dependence of the nuclear path length in connection with the microscopic
in-medium width in the propagator. This leads to a slight improvement in
the description of the invariant-mass spectra measured by the CLAS
Collaboration for iron targets, which we also extended to carbon
targets. The sensitivity to medium
effects in the inclusive mass spectra is, however, somewhat limited in
the absence of absolutely normalized data.
Calculations for heavier target nuclei indicate that the situation
improves when increasing the nuclear-mass number to $A\simeq200$. A more powerful lever arm
turns out to be a cut on the outgoing lepton pair momentum. For small
momenta, say, $q\le 0.5$~GeV, the spectral shape becomes very flat
with hardly any $\rho$ resonance peak left. The reason for this
effect is twofold: on the one hand the kinematics make slow $\rho$ mesons stay
longer in the nuclear volume and thus enhance the in-medium
decay fraction; but dynamics
cause the medium effects in the ρ spectral function to increase
substantially toward small three-momentum. It is this dynamic effect that reconciles
the large medium effect observed in heavy-ion collision (average $\rho$
width of $\sim 350$--$400$~MeV~\cite{vanHees:2007th} as extracted from
the NA60 data on
low-mass dimuons) with the rather moderate modifications
($\Gamma_\rho\simeq 220$~MeV from the current CLAS data) in nuclear
photoproduction. This clearly calls for a low-momentum measurement in
the latter experiments, to scrutinize the 3-momentum dependence of the
medium effects as well as the prevalence of baryon-induced modifications
in the interpretation of the dilepton data in ultrarelativistic heavy-ion
collisions.
We furthermore found that the nuclear dependence of absolute production
cross sections can be used as a rather sensitive, complementary measure
for the in-medium width. These are conveniently analyzed using
the so-called nuclear transparency ratio. While the sensitivity to
an increase in the absolute value of the width is less pronounced
for the $\rho$ than for the narrow vector mesons $\omega$ and $\phi$,
we still predict a substantial reduction of this ratio, which should be
straightforward to measure. An important point here is that, for the
$\rho$, the yield measurements can be effectively combined with, and
tested against, the spectral information from the invariant-mass
distributions (which is very challenging for the narrow vector mesons
owing to their large decay fraction outside the nuclear volume).

As a further direct extension of the present work, we plan to study
dilepton spectra in nuclear electroproduction
(via the reaction $ e^-\,A\,\rightarrow e^- \,e^+e^- A$), which
would provide further constraints by investigating the various
mechanisms discussed in this article under modified kinematics.
It would also be interesting to study $\pi^+\pi^-$ final states,
which are more easily accessible experimentally~\cite{Morrow:2008ek}
but suffer from extra absorption factors on the outgoing pions
thus reducing the sensitivity to the central densities in nuclei.
It would be very illuminating to perform detailed comparisons
to transport theoretical calculations. The latter, in principle,
performs a better treatment of the off-equilibrium aspects in the
propagation process, whereas the implementation of broad spectral
functions is more involved than in equilibrium approaches like
the one adopted here.
A continued effort on both experimental and theoretical fronts,
in both heavy-ion and nuclear production contexts, holds exciting
prospects for further progress in the understanding of hadron
properties in the medium.

\vspace{0.5cm}


\acknowledgments

We are grateful to C.~Djalali for fruitful discussions.
F.R. and R.R. were supported by the U.S. National Science Foundation
through a CAREER grant No. PHY-0449489.
T.-S.H.L. was supported by the U.S. Department of Energy,
Office of Nuclear Physics Division, under contract No. DE-AC02-06CH11357.


\end{document}